\documentclass[letterpaper]{emulateapj} 
\usepackage{apjfonts}                   

\usepackage{epsfig,graphicx,latexsym,amsmath,amssymb}
\usepackage{natbib}
\usepackage{hyperref}
\usepackage{mathrsfs}
\usepackage{lastpage}

\bibpunct[,]{(}{)}{;}{a}{}{,}

\newcommand{\Mcl}{\ensuremath{{\cal M}_{\rm cl}}}
\newcommand{\MBH}{\ensuremath{{\cal M}_{\bullet}}}
\newcommand{\Rh}{\ensuremath{R_{\rm h}}}
\newcommand{\kms}{\ensuremath{\,\mathrm{km\,s}^{-1}}}
\newcommand{\msol}{\ensuremath{{\rm M}_{\odot}}}
\newcommand{\Msun}{\msol}

\newcommand{\Vinfty}{\ensuremath{V_{\infty}}}
\newcommand{\Myr}{\ensuremath{\rm Myr}}

\begin{document}

\author{
Pau Amaro-Seoane\altaffilmark{1}\thanks{e-mail: Pau.Amaro-Seoane@aei.mpg.de},
Symeon Konstantinidis\altaffilmark{2}\thanks{e-mail: simos@ari.uni-heidelberg.de},
Marc Dewi Freitag\altaffilmark{3,\,4}\thanks{e-mail: marc.freitag@gmail.com},
M.\ Coleman Miller\altaffilmark{5}\thanks{miller@astro.umd.edu},
\& Frederic A.\ Rasio\altaffilmark{6}\thanks{e-mail: rasio@northwestern.edu}
}

\altaffiltext{1}{Max Planck Institut f\"ur Gravitationsphysik
(Albert-Einstein-Institut), D-14476 Potsdam, Germany}
\altaffiltext{2}{Astronomisches Rechen-Institut, M{\"o}nchhofstra{\ss}e 12-14, 69120,
Zentrum f\"ur Astronomie, Universit\"at Heidelberg, Germany}
\altaffiltext{3}{Institute of Astronomy, Madingley Road, CB3 0HA Cambridge, UK}
\altaffiltext{4}{Gymnase de Nyon, Route de Divonne 8, 1260 Nyon, Switzerland}
\altaffiltext{5}{Department of Astronomy and Joint Space-Science Institute,
University of Maryland, College Park, MD 20742-2421, USA}
\altaffiltext{6}{Department of Physics and Astronomy, and Center for
Interdisciplinary Exploration and Research in Astrophysics (CIERA),Northwestern University, Evanston,
IL 60208, USA}

\date{\today}

\label{firstpage}

\title{Sowing the seeds of massive black holes
       in small galaxies:\\
       Young clusters as the building blocks of Ultra-Compact-Dwarf Galaxies}

\begin{abstract}
Interacting galaxies often have complexes of hundreds of young stellar clusters
of individual masses $\sim 10^{4-6}~M_\odot$ in regions that are a few hundred
parsecs across.  These cluster complexes interact dynamically, and their
coalescence is a candidate for the origin of some ultracompact dwarf galaxies
(UCDs).  Individual clusters with short relaxation times are candidates for the
production of intermediate-mass black holes of a few hundred solar masses, via
runaway stellar collisions prior to the first supernovae in a cluster.  It is
therefore possible that a cluster complex hosts multiple intermediate-mass
black holes that may be ejected from their individual clusters due to mergers
or binary processes, but bound to the complex as a whole.  Here we explore the
dynamical interaction between initially free-flying massive black holes and
clusters in an evolving cluster complex.  We find that, after hitting some
clusters, it is plausible that the massive black hole will be captured in an
ultracompact dwarf forming near the center of the complex.  In the process, the
hole typically triggers electromagnetic flares via stellar disruptions, and is
also likely to be a prominent source of gravitational radiation for the
advanced ground-based detectors LIGO and VIRGO.  We also discuss other
implications of this scenario, notably that the central black hole could be
considerably larger than expected in other formation scenarios for ultracompact
dwarfs.
\end{abstract}

\maketitle

\section{Introduction}

Several bound
systems of young, massive clusters in colliding galaxies have been
observed using the Hubble Space Telescope (HST).  The best studied
case is the Antenn{\ae} galaxies (NGC 4038/4039), the nearest example of two
colliding disc galaxies listed in the \cite{Toomre_list} sequence. HST
observations reveal in this system the existence of relatively small regions
(compared with the size of the galaxies) harbouring hundreds or thousands of
young clusters \citep{HST_Antenae_Revisited_2010, HST_Antenae_1999,
Cluster_Complexes_review}. In particular, \cite{HST_Antenae_Revisited_2010}
observed $18$ areas (``knots'') of sizes spanning $100-600$~pc which
contain hundreds of clusters. The mass function of those systems, which we will
henceforth refer to as ``Cluster Complexes'' (CCs), is

\begin{equation}
dN/dM \propto M^{\beta},
\label{eq.IMFCC}
\end{equation}

\noindent
with $\beta = -2.10 \pm 0.20$ in the range $M\sim 10^{4-5}~M_\odot$ \citep[see
also][]{CC_Mass_function}.
\cite{Low_mass_CCs_Antennae} found in the same system low-mass CCs with masses
around $10^6 \rm M_\odot$ and diameters of some $100 - 200$~pc. One of the
best studied CCs in the Antenn{\ae} galaxy is ``knot S'', with a total mass of
$10^8 \rm M_\odot$ and a total radius of $\sim 450 \rm pc$
\citep{HST_Antenae_1999}.  Other galaxies with recently discovered CCs
include NGC~7673 \citep{CCs_in_NGG7683},
M82 \citep{CCs_in_M82}, NGC~6745 \citep{deGrijsEtAl03},
Stephan's Quintet \citep{CCs_In_Stephans_Quintet} and NGC~922
\citep{CCs_in_NGC922}.

CCs are bound systems
\citep{Kroupa1998,Kroupa2005,Kroupa2011,HST_Antenae_Revisited_2010} and on
relatively short time-scales at least some of their member clusters will merge
to form a single object. \cite{Kroupa1998} and \cite{Kroupa2005} have
postulated CCs as the breeding ground of Ultra-Compact Dwarf Galaxies (UCDs).
Following this idea, \cite{Kroupa2011} performed $N-$body simulations of CCs
with different total masses
($10^{5.5} - 10^{8}~M_\odot$) and initial Plummer radii $10-160$~pc.
They conclude that UCDs, Extended Clusters (ECs) or even large Globular
Clusters (GCs) might be the product of an agglomeration of clusters in CCs.
They find in their simulations that almost all members of a CC merge in
less than 1~Gyr. In some cases this timescale can be as short as 10~Myr.
By the end of their simulations a very massive cluster forms in the
centre of the CC, with a mass of $26-97\%$ the mass of the initial CC and a
radius of $\sim 50$~pc.

Theoretical and numerical studies show that at least a fraction of young star
clusters could host intermediate-mass black holes (IMBHs, black holes with
masses ranging between $10^{2-4}\,M_{\odot}$) at their centres. A possible
formation path is that in a young cluster, the most massive stars sink to the
centre due to mass segregation.  After a high-density stellar region forms,
stars start to collide and merge with each other. A number of numerical studies
with rather different approaches show that, under these circumstances, at least
one of the stars increases its mass rapidly in a process of runaway collisions
\citep{PortegiesZwartMcMillan00,GurkanEtAl04,PortegiesZwartEtAl04,
FreitagEtAl06,FreitagRB06}.  Nonetheless, there are a number of open questions
regarding this process.  One of the main uncertainties is the role of
stellar winds.  In principle at approximately solar metallicity winds may limit the mass
of this very massive star (VMS) to a few hundreds of solar masses rather than a
few thousands \citep{BelkusEtAl07}. Nevertheless we note that this requires a
substantial extrapolation of already uncertain wind loss rates to stellar
masses an order of magnitude beyond what is observed. Also, the collision
process might lead to lumpy bags of stellar cores in an extended envelope rather
than to relaxed stars near the end of the runaway collision (M. Davies, private
communication).  In addition, when
\cite{SuzukiEtAl07} combined direct $N-$body simulations with smooth particle
hydrodynamics (SPH) they found that stellar winds would not hinder the formation
of the VMS.  It is thus possible but not certain that IMBHs can form in
young clusters.  We will assume their existence as a working hypothesis.

Apart from the obvious interesting implications for models of galaxy formation
and, in particular, of UCDs, mergers of clusters in CCs are a powerful source
of gravitational waves if these harbour central IMBHs in their respective
centres \citep{pau2006,AS10a,Amaro-SeoaneEtAl09a,Amaro-SeoaneSantamaria10}. In
particular, \cite{pau2006} showed that such mergers would lead to the formation
of an IMBH binary, which would merge in a time scale as short as $\sim 7$~Myr.
Such a merger would be easily detected with space-borne observatories and also
with ground-based detectors such as Advanced LIGO or Advanced VIRGO
(AdLIGO/AdVIRGO) if it occurs within $\sim 2$~Gpc \citep{FregeauEtAl06}. Using
more realistic waveforms including spins, \cite{Amaro-SeoaneSantamaria10} find
that the detection distance is increased significantly, up to an
orientation-averaged distance of $\sim 5 - 12$ Gpc, depending on the spin
configuration and mass ratios. In the case of the Einstein Telescope (ET), the
same authors find that the maximum redshifts for ET are $z \sim 10$, which
implies that binaries of IMBHs will be a cosmological probe.

Numerical relativity simulations show that during the merger of the holes,
gravitational radiation is emitted asymmetrically with the size of asymmetry
depending on the mass ratio of the two black holes and on their spin magnitude
and orientation \citep{Recoil_review,Recoil0,Recoil1,BakerEtAl08,LoustoEtAl10,Recoil4,Recoil5,HealyEtAl09,BinaryBH,Recoil_without_spins,HerrmannEtAl07a,HerrmannEtAl07b,BoyleKesden08,vanMeterEtAl10,LoustoZlochower11a}
If this recoiling velocity exceeds a few times the velocity dispersion of the
merged cluster, then the IMBH leaves the host cluster.  There is a massive
black hole at large in the CC.  Even if an IMBH escapes from one cluster, it
might still be bound to the CC as a whole, which means that it has the
possibility of interacting with other clusters and, perhaps, their IMBHs.

In this article we address the formation of ultra-compact dwarf galaxies by the
agglomeration of young clusters in CCs, along with the role of one or more
recoiling IMBHs, using direct-summation $N-$body simulations. For this, we run
a set of $\sim\,200$ individual experiments in which we vary mass ratios,
relative speeds, and impact parameter to study in detail the interaction
between a single IMBH and a cluster.  We then study the interaction of one or
more IMBHs at large in a CC with individual clusters with an additional set of
$N-$body simulations.  We correct for the trajectory of the IMBH, based on
point dynamics and the mass loss in the individual clusters, by using the
previous 200 experiments.  We also follow the growth of a seed UCD in a CC and
record all stellar disruptions triggered by the presence of the IMBH(s). For
realistic models of CCs we find that the IMBH(s) end up captured by the seed
UCD or by a smaller cluster which is close to the UCD. Thus, if the fraction of
IMBHs in the CC ($f_{\bullet}$ from now onwards) is not zero, this is a process
of allocating one or more IMBH at the very centre of a UCD.

\section{Interactions between a recoiling IMBH and an individual young cluster}
\label{sec.interactions_MBH_Cl}

\begin{figure}
\resizebox{\hsize}{!}{\includegraphics[scale=1,clip]{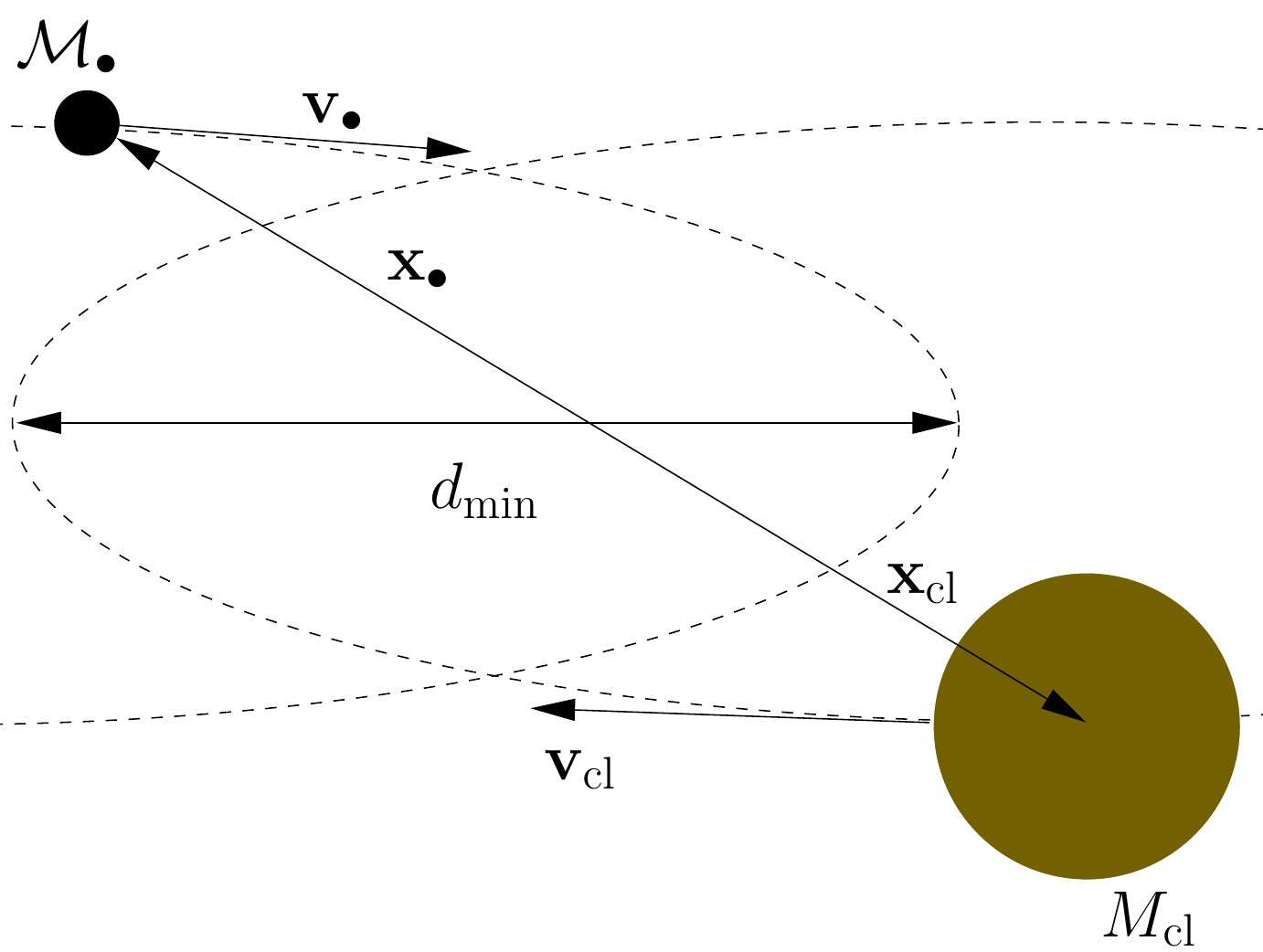}}
\caption{
  Geometry for the initial conditions of the parabolic collision, in the COM of
  the IMBH--cluster system. To obtain the grid displayed in
  Fig.(\ref{fig.outcomes}), we systematically vary $d_{\rm min}$, the
  relative velocity and the mass ratio between the IMBH and the cluster.
\label{fig.parabola}
}
\end{figure}

In this section we make a study of the parameter space for a collision between
a recoiling IMBH and an individual young cluster in a CC. We run a set of
$\sim$ 200 direct $N-$body simulations to build a grid which we will later use
in our simulations of the IMBH in the CC, as explained in the introduction.
Initially we set the IMBH and the cluster on an orbit with positive relative
speed and thus positive total energy in the initial state, i.e. a hyperbolic
orbit, as described in \cite{Amaro-Seoane06}. We schematically show this for
reference in Fig.\,(\ref{fig.parabola}) and follow a similar notation.  The
initial trajectory of the IMBH would bring it within a minimum distance $d_{\rm
min}$ of the cluster centre if the cluster was replaced by a point particle. In
the centre-of-mass reference frame (COM),

\begin{align}
{\bf x}_{\bullet} & =  \lambda_{\rm cl} \,{\bf d},       \nonumber \\
{\bf x}_{\rm cl}  & = -\lambda_{\bullet} \,{\bf d},      \nonumber \\
{\bf v}_{\bullet} & =  \lambda_{\rm cl} \,v_{\rm rel},   \nonumber \\
{\bf v}_{\rm cl}  & = -\lambda_{\bullet} \,{\bf v}_{\rm rel}
\end{align}

\noindent
where ${\bf v}_{\rm rel}$ is the relative velocity of the two objects
${\bf d}$ is their separation vector, ${\bf
x}_{\bullet\,,{\rm cl}}$ are the positions of their centres,
and $\lambda_{\bullet\,,{\rm cl}}={m_{\bullet\,,{\rm
cl}}}/({{\cal M}_{\bullet}+{M}_{\rm cl}})$.

\noindent

The number of stars in the cluster is always ${\rm N}_{\star} =3\times 10^{4}$
and we use for their initial distribution a King model of concentration $W_0=7$
\citep{King66,PetersonKing75}, and all stars have the same mass, to simplify
the interpretation of the results, although we note that a mass function could
have an impact in the outcome of individual hits. Stellar evolution is not
taken into account for the same reason. Although the number of stars we
simulate is still below of what we can expect from a real cluster, we deem the
dynamical interaction to be correct but for probably the most extreme mass
ratios in which the mass ratio between the IMBH and the total mass in the
cluster is one and two. We include these cases for completness but note that in
those cases the stars in those clusters do not represent a single star but a
set of them. I.e. the IMBH will hit lighter clusters with those mass ratios,
and the orbital evolution of the IMBH will be correctly estimated in our
simulations, but the trajectory of a single star in such clusters does not
trace one single star, but a set of them.  The simulations are performed with
the direct-summation NBODY4 programme of \cite{Aarseth03}. This choice was made
for the sake of the accuracy of the study of the orbital parameter evolution of
the IMBH and mass loss in the cluster; this numerical tool includes both the KS
regularisation \citep{KS65} and chain regularisation, which means that when two
or more particles are tightly bound to each other or the separation is very
small during a hyperbolic encounter, the system becomes a candidate to be
regularised in order to avoid problematical small individual time steps.  The
basis of direct $N-$body codes relies on a Hermite integrator scheme
\citep{Aarseth99,Aarseth03} for which we need not only the accelerations but
also their time derivatives. This extra computational overhead is necessary for
us to follow reliably the orbital evolution of {\em every} single star (or
IMBH) in our system.
While the code was not meant to integrate clusters in which a particle is significantly
much more massive than the rest of them, a mass ratio of the order of what we have considered
in this study leads to an accurate integration, with individual time integration errors of the
order of $10^{-10}$ in energy.

At the end of an $N$-body run, we need to identify the particles that are still
forming a bound cluster, the particles that are bound to the IMBH, and the
particles that have become unbound.  We also need to know whether the IMBH has
been captured by the stellar cluster. We have therefore developed an
iterative algorithm.  To initialize the procedure, we make a (computationally)
quick guess of which particles are bound to the cluster and which ones form a
bound group including the IMBH (the ``IMBH group''). Note that a given particle
can be in both groups, for instance if the IMBH has been captured by the cluster
and has sunk to its centre or is orbiting it. For this first guess, stellar
particles are considered bound to the IMBH group if they are bound to the IMBH
(i.e., we do not take into account the self-gravity of the bound stars
themselves).

For the first-guess cluster, one assumes that its centre corresponds to the
median position of all stellar particles, i.e. the $x$, $y$  and $z$ components
of the ``centre'' are taken to be the medians of the corresponding components
of the positions of all the stellar particles.  The median turns out to be a
much more robust estimate of where the bulk of the particles is, compared to
the average position or the centre of mass (i.e. the mass-weighted average
position) as those quantities are very sensitive to the the positions of a few
particles ejected at large distances from the rest.  For this first guess, the
90\,\% of the particles closest to this median position are assumed to be part of the
cluster.

\begin{figure*}
\resizebox{\hsize}{!}{\includegraphics[scale=1,clip]{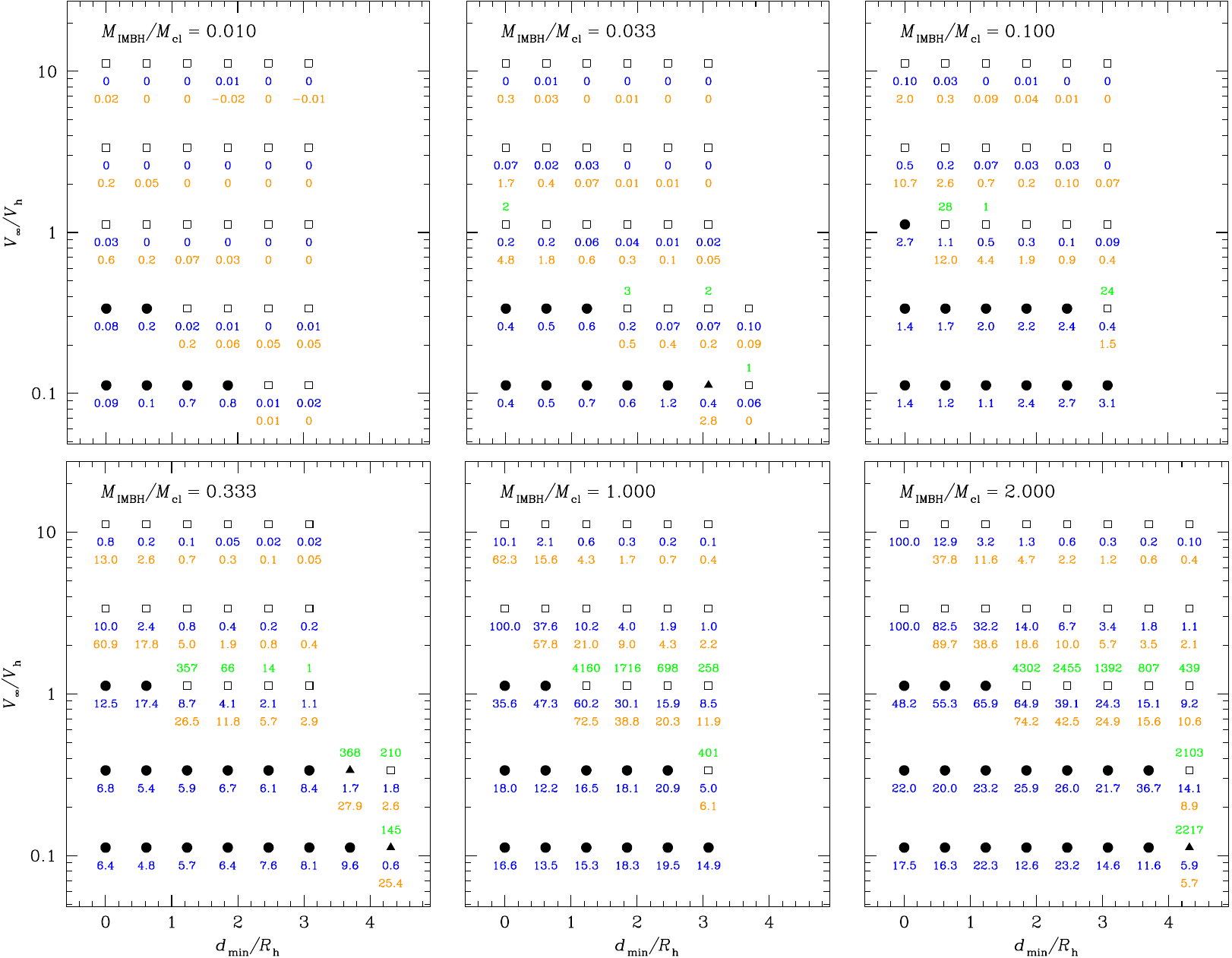}}
\caption{
Outcomes of all 196 simulations of encounters between a cluster with King
parameter $W_0=7$ and an IMBH.  Each panel shows the results for a given mass
ratio $\MBH/\Mcl$. The abscissa of each plot is the minimum distance $d_{\rm min}$,
computed assuming 2-body dynamics, in units of the half-mass radius $\Rh$. The
ordinate is the relative velocity at infinity $\Vinfty$, in units of $V_{\rm
h}\equiv (G\Mcl/\Rh)^{1/2}$, a typical velocity dispersion for the
cluster. Solid round dots show ``mergers'', i.e., cases where the IMBH has been
captured by the cluster and has settled at its centre. Solid triangles are
cases in which the IMBH is orbiting the cluster (a merger is likely to be the
long-term outcome). Open squares are ``fly-throughs''. The number just below a
symbol (in blue in the on-line colour version) is fractional mass loss from the
cluster in percent. The second, lower number (in orange in the on-line colour
version) is the fractional reduction in specific binding energy
of the cluster, also in percent. A
number above a symbol indicates how many stellar particles are bound to the IMBH
(when it has not merged with the cluster).
\label{fig.outcomes}
}
\end{figure*}

For the first iteration, we have to compute the binding energy of a particle
relative to the cluster group, hence we need to know the velocity of that
group. To estimate the velocity of the cluster in the first-guess attribution,
we take the average velocity of the 10\,\% of the particles closest to the assumed
centre. This number is sufficient to avoid large fluctuations due to
individual particle velocities (``random velocities'').  On the other hand,
taking a significantly larger fraction of particles is neither necessary nor
advisable as it is not yet known which particles are actually bound together as
a cluster. We have to make sure that the velocity defined in the procedure is a
good estimate of that of the actual bound cluster. Otherwise, the kinetic
energies relative to this first-guess cluster are biased towards high values
and the iterative procedure fails at identifying a bound cluster.  The
iterations proceed as follows: For each particle, the binding energies relative
to the cluster and the IMBH group are computed. For this, we have to estimate
the position of the centre of each group and its velocity. For the IMBH
group, they are fixed to the values of the IMBH itself. For the cluster, the
centre position and velocity are defined to be the mass-weighted mean
values for all particles within half a ``typical size'' of the previous
estimate of the centre. The typical size of the cluster is the harmonic mean
of the distance to its centre (for all particles considered bound to it):

\begin{equation}
R_{\rm typ} = R_{\rm harm} \equiv M_{\rm cl}\left(\sum\frac{m_i}{R_i}\right)^{-1}.
\end{equation}

\noindent
One advantage of defining $R_{\rm typ}$ using the harmonic mean, instead of using the
half-mass radius or some other Lagrangian radius, is that this does not
require a sorting of the particles.  The gravitational energy is computed
assuming a spherical mass distribution, i.e., as if each particle bound to a
group (cluster or IMBH group) was a spherical shell of matter, of radius $R_i$
centred on either the IMBH position or the estimated centre of the cluster.
Typically, the attributions of the particles to either or both groups
converge after fewer than ten iterations.

\noindent
At the end, the attributions are cleaned up in the following way. If a stellar
particle belongs to both the cluster and the IMBH group, the binding energies
to both structures are compared. It will be kept as member of the IMBH group only
if the binding energy to the IMBH group is larger than to the cluster group.
In that case, it will also be
kept as member of the cluster only if the IMBH
itself is bound to the cluster. This reduces the number of double-members in a
reasonable way, still allowing for situations such as the IMBH having captured
some stars while being itself on a bound orbit around the (main) cluster.

\noindent
Finally, to interpret the results, we allow for three different outcomes.  A {\it
merger} is when the IMBH group is bound to the cluster (as determined assuming
each group is a point mass) and the distance between the centres of the groups
is smaller than the sum of the $R_{\rm typ}$'s. A {\it satellite} situation
arises when the two groups are bound but the distance between their centre is
larger than twice the sum of the $R_{\rm typ}$'s. A {\it flyby} is when the
groups are unbound and the distance between their centres is larger than either
the sum of the total extent of each group or five times the sum of the $R_{\rm
typ}$'s. Any other situation would be considered as {\it unknown} but does not
occur if the $N-$body simulation has been carried out for a sufficient duration.

\begin{figure*}
\resizebox{\hsize}{!}
          {\includegraphics[scale=1,clip]{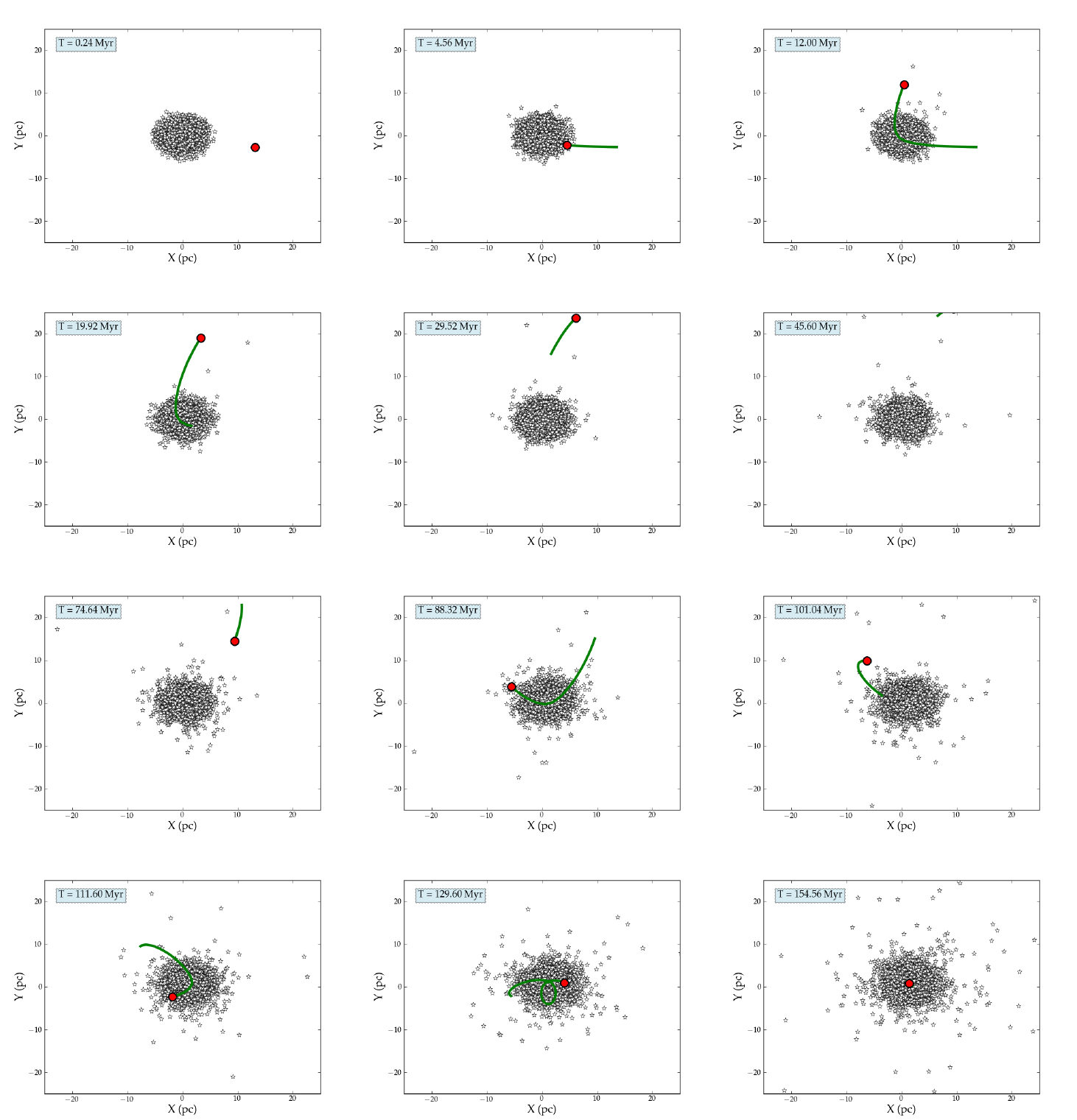}}
\caption
   {
Projection in the X--Y plane of all trajectories of the stars (star symbols) in
a cluster and the IMBH (red circle) for 12 different moments in the interaction.
In this particular case, the process leads to the capture of the IMBH. For
visibility, the radius of the IMBH and the stars has been artificially
magnified.  We also depict the previous 60 positions of the IMBH with a
solid, green line. The mass ratio between the IMBH and the cluster is 0.01,
the minimum distance of approach of the COM of the cluster and the IMBH is
$d_{\rm min} = 1$ and $\Vinfty = 1\kms$.
   }
\label{fig.dmin1_IMBH_300_Vnfty_1kms}
\end{figure*}

\begin{figure*}
\resizebox{\hsize}{!}
          {\includegraphics[scale=1,clip]{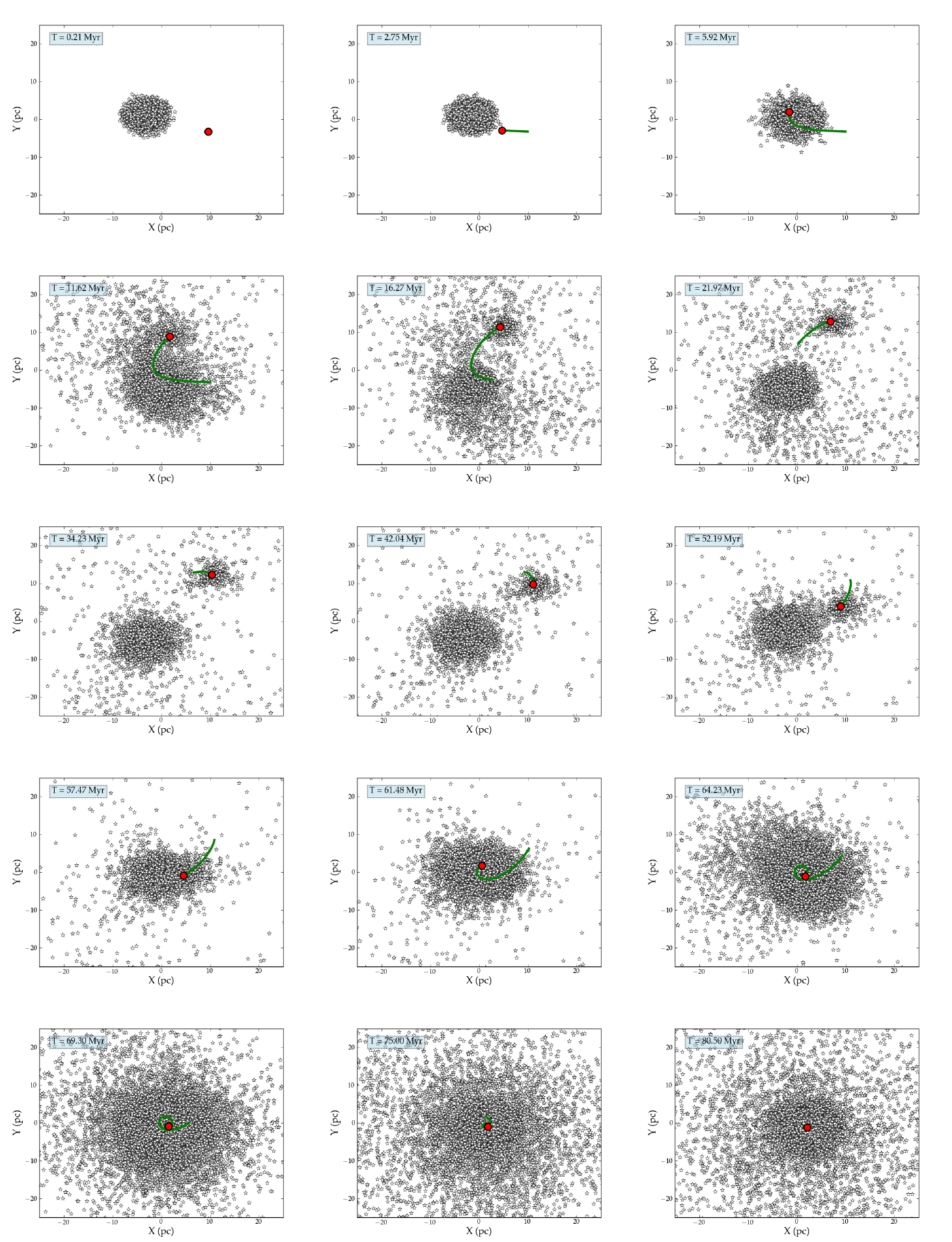}}
\caption
   {
Same as in Fig.(\ref{fig.dmin1_IMBH_300_Vnfty_1kms}) for 15 different times.
The mass ratio in this case is 0.333, $d_{\rm min} = 5$ and $\Vinfty = 3\kms$.
}
\label{fig.dmin5_IMBH_10k_Vnfty_3kms}
\end{figure*}

\noindent
In Fig.(\ref{fig.dmin1_IMBH_300_Vnfty_1kms}) and
\ref{fig.dmin5_IMBH_10k_Vnfty_3kms} we show two particular cases for the
IMBH -- cluster interaction in the COM frame which, although not representative
for the whole sample displayed in Fig.(\ref{fig.outcomes}), are interesting in
terms of the dynamics of the system
\footnote
        {
The interested reader can visit\\
\url{http://members.aei.mpg.de/amaro-seoane/ultra-compact-dwarf-galaxies},
\\
for movies based
on the results of the figures (the last URL is a 3D version of the second movie).
The encoding of the movies is the free OGG Theora format
and should stream automatically with a gecko-based browser (such as mozilla or firefox)
or with chromium or opera.  Otherwise please see e.g.\\
\url{http://en.wikipedia.org/wiki/Wikipedia:Media_help_(Ogg)}\\
for an explanation on how to play the movies.
        }
. In the first case $d_{\rm min} = 1$, which
leads to an almost head-on collision between the IMBH and the cluster. Nonetheless,
because of the low relative velocity and mass ratio, the interaction does not
lead to a huge mass loss from the cluster. Even if at $T = 45.60$ Myr the IMBH and
cluster seen to be unbound, the IMBH is still forming a binary with the COM of
the cluster and, hence, the semi-major axis decays again. After some 154 Myrs
the IMBH settles down to the centre and is captured. In the second figure, the
larger mass ratio has a significant impact in terms of mass loss. Already after
11.62 Myr the IMBH has captured some stars from the cluster, which remain
bound to the trajectory of the hole and follow its trajectory. This satellite
and the IMBH are nevertheless still gravitationally bound to the cluster and
hence fall back again. The higher mass in the IMBH--satellite system leads to
a rather large mass loss from the original cluster. After 80.50 Myrs the IMBH
is at the centre of the remaining cluster.

\section{Interactions between a single recoiling IMBH and clusters in a CC}
\label{sec.larger_scale}

\subsection{Integrator and first considerations}

Now that we have completed the grid of individual IMBH-cluster interactions, we
can explore the scenario in which one IMBH is at large in a CC, interacting with
many different IMBH on its way, either to an eventual escape from the CC, or down
to the very centre, where the seed of a UCD is forming by the mergers
of clusters. In this section we will
assume $f_{\bullet} = 2/N_{\rm tot}$, where $N_{\rm tot}$ is the total number of
clusters in the CC; that is, only two clusters in the whole CC harbour an
IMBH and we also assume that they have coalesced and the merged hole escaped
from the host cluster. As we will see, the presence of the IMBH triggers stellar
disruptions in individual clusters of the CC, which could potentially represent
a fingerprint of this process. In next section we will
look at larger values of $f_{\bullet}$.

The numerical code that we use for the simulations of the CC and the IMBH is
\texttt{Myriad} \citep{Myriad}, which uses the Hermite fourth-order
predictor-corrector scheme with block time steps \citep{Hermite4} for advancing
the particles in time, while the accelerations and their derivatives are
computed using GRAPE-6 \citep{Grape6} special purpose computers.  Close
encounters between particles (i.e. between clusters or between the IMBH and a
cluster) are detected using the GRAPE-6 and evolved with a time-symmetric
Hermite fourth-order integrator \citep{tsHermite4}.  Even though the code was
originally designed for dynamical simulations of stars in star clusters, its
flexible modularity made it easy to adapt to our particular problem. In
particular, we assigned a radius to each particle representing a cluster, and
we allowed clusters to merge with each other whenever the distance was smaller
than the sum of the radii.  In the simulations the IMBH is also a particle with
a radius set to its Schwarzschild radius.

From the individual IMBH-cluster simulations presented previously we have data
for the outcomes based on the mass ratio ${\cal M}_{\bullet}/\Mcl$, the distance
of closest approach between IMBH and cluster, and the relative velocity of the two
objects and, thus, the change in kinetic energy of the IMBH. We use these
results to correct the position and velocity of the IMBH after each interaction
with a cluster in the simulation of the CC. This also provides us with
information about the number of stellar disruptions triggered by the IMBH, as
well as the characteristics of the cluster which captures the IMBH (if any).  If
a capture does occur, the simulation finishes and then we record the position of
the ``trapping'' cluster in the CC. Another possible termination of the
simulation is if the IMBH leaves the CC, because its speed is high enough to
escape the complex.

\subsection{Assumptions for the initial conditions of the CC and the IMBH}

Initially we fix the radius of the CC, $R_{\rm CC}$, to a typical value coming
from observational data and populate it with individual clusters following
equation \ref{eq.IMFCC}. In particular, in the ``knots'' of the Antenn{\ae}
galaxy one observes a mass distribution with $n=-2$. The number of {\em observed} individual
clusters in CCs is of the order of 100, but
the actual number might actually be thousands, most of which are simply too
faint to be observed \citep[as discussed in e.g.][]{Kroupa2005}. We set the
total mass of the CC to a typical observed value, $M_{\rm CC} \sim
10^6-10^8 \, M_\odot$.  The individual clusters have half-mass radii ranging
between 0.5 and 4 pc and are distributed initially in the CC following a
Plummer model \citep{Plummer} with a cut off radius (see table
\ref{tab.PowerLaw}). The masses of the clusters are discrete and come from the
$\rm {\cal M}_{\bullet}/\Mcl$ ratios that were used in the IMBH--cluster
$N-$body simulations. Then, for ${{\cal M}_{\bullet}} = 5\times 10^3 {M_\odot}$
and $\rm {\cal M}_{\bullet}/\Mcl = 0.01, 0.033, 0.1, 0.33, 1, 2$, the discrete
masses of the clusters in the CC are $5 \times 10^5\rm{M_\odot}, 1.51515 \times
10^5 \rm{M_\odot}, 5 \times 10^4\rm{M_\odot}, 1.51515 \times 10^4\rm{M_\odot},
5\times 10^3\rm{M_\odot} \, \rm{and} \, 2.5\times 10^3 \rm{M_\odot}$. When
two clusters collide in the CC simulation,
we assume a 20\% mass loss, based on the simulations of the collisions of two clusters of \cite{ASF06,Amaro-SeoaneEtAl09a}, so the
cluster product of the merger of two individual clusters has a mass which is
80\% of the sum of the masses and a new radius, equal to the radius of the
more massive cluster plus the 20\% of the sum of the radii of the two clusters.

\begin{figure}
\resizebox{\hsize}{!}
          {\includegraphics[scale=1,clip]{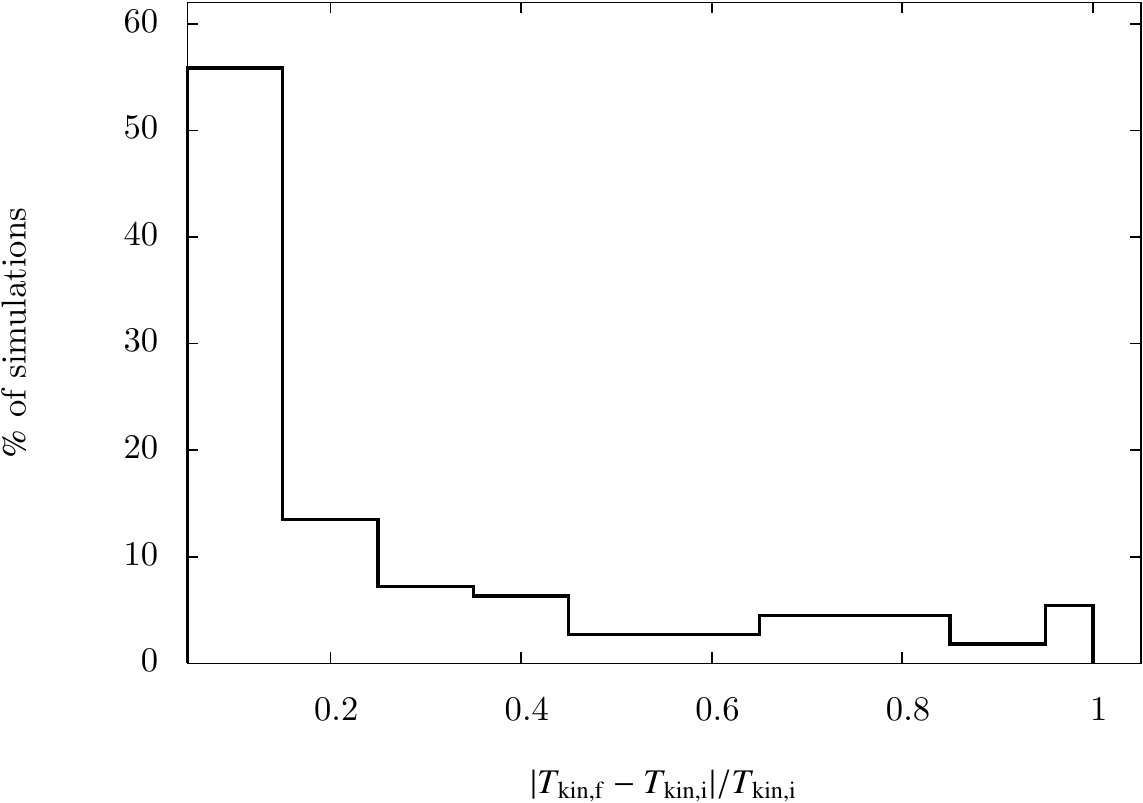}}
\caption
   {
Kinetic energy difference between the initial and final kinetic energy ($T_{\rm kinf,\,i}$ and $T_{\rm kinf,\,f}$
respectively) normalized to the initial energy for all
collisions between the IMBH and the clusters resulting in a fly-through for all $N-$body simulations.
   }
\label{fig.histogram}
\end{figure}

The IMBH in the merged cluster is assumed to be the product of a merger of two
IMBHs that were located at the centres of two merging star clusters.  We assume
that this happened close to the centre of the CC, where most of the individual
cluster-cluster collisions take place, because this is where the numerical
density of clusters is highest.  Hence, we initially place the IMBH at the
centre of the CC. We choose a mass of ${{\cal M}_{\bullet}} = 5\times
10^3~M_\odot$, which determines the masses of individual clusters from the grid
given in the previous section.  The recoil speed of the merged IMBH could in
principle be up to $\sim 5000$~km~s$^{-1}$
\citep{HerrmannEtAl07a,HerrmannEtAl07b,BoyleKesden08,LoustoZlochower11a,LoustoZlochower11b}
for optimal mass ratios, spins, and spin orientations.  The recoils of greatest
interest to our present study are in the $\sim 100$~km~s$^{-1}$ range, because
the merged IMBH will then escape from its host cluster but be bound to the CC
as a whole.  It is difficult to judge how representative this will be for the
mergers of actual IMBHs in CCs.  Assuming the spin orientations are random,
speeds in this range are characteristic of mass ratios $q\sim 0.1$ for
substantial spins, or spins $a/M\sim 0.1$ for mass ratios comparable to unity
\citep{Recoil_stats,vanMeterEtAl10}.  For our purposes we will study the case
of $v_{\rm recoil}=100$~km~s$^{-1}$.  At this speed, the escape time from a
cluster of total radius $\sim 10 \rm pc$ is $\sim 0.1 \rm Myr$. Hence we simply
place the IMBH initially at the centre of the CC, not bound to any cluster, and
assume that it recoils in a random direction.

For the evolution of the recoiling IMBH we must take into account the loss of
kinetic energy every time it hits a cluster. In figure (\ref{fig.histogram}) we
can see the distribution of the resulting kinetic energy after a hit for all
fly-by simulations of figure (\ref{fig.outcomes}).  While there is a spread in
the distribution, there is a strong spike around 10\% of loss for about 50\% of
all simulations.  We have therefore adopted a slightly larger value, of 20\%.
This loss of energy will result into a rather negligible deacceleration of the
IMBH, so that it will have more chances to escape the CC, and it will also lead
to a lower number of stellar tidal disruptions.  On the other hand, a bit less
than 50\% of all ``fly-throughs'' have {\em at least} over $\sim 5\%$ of
relative loss after one hit. This situation is more appealing from a dynamical
standpoint, and therefore we will first address it.  In the next sections we
will assume an average loss of $20\%$ for the ``fly-throughs'' hits, and in
section \ref{sec.5percent} we will briefly explore the other situation.

\begin{table*}
\begin{center}
\begin{tabular}{  l | c | c | c | c | l | c | c | c }
\hline
  ID & $ N $ & $M_{\rm CC}\, (M_\odot)$ & $R_{\rm CC}$ (pc)& $ $ & $ {\rm ID} $ & $ N $ & $M_{\rm CC}\, (M_\odot)$ & $R_{\rm CC}$ (pc) \\
  \hline \hline \hline
   A1  & $ 5 \times 10^2 $ & $ 1.522 \times 10^7$ & $ 45$ & $ $ &  E1  & $ 3\times 10^3 $ & $ 4.32 \times 10^7$ & $122$  \\
  \hline
   A2  & $ 5 \times 10^2 $ & $ 1.522 \times 10^7$ & $ 90$ &  $ $ &  E2  & $ 4 \times 10^{3} $ & $5.75 \times 10^7 $ &$165$\\
  \hline
   A3  & $ 5 \times 10^2 $ & $ 1.522 \times 10^7$ & $ 132$ &  $ $ &  E3  & $ 4 \times 10^{3} $ & $5.75 \times 10^7$ & $246$\\
  \hline
   A4  & $ 5 \times 10^2 $ & $ 1.522 \times 10^7$ & $ 168$ &  $ $ &  E4  & $ 4 \times 10^{3} $ & $5.75 \times 10^7$ & $329$\\
  \hline
   A5  & $ 5 \times 10^2 $ & $ 1.522 \times 10^7$ & $ 255$ &  $ $ & $  $ & $ $ & $ $ & $ $\\
  \hline \hline
   B1  & $ 1 \times 10^3 $ & $ 1.522 \times 10^7$ & $ 90$ & $ $ &  F1  & $ 5 \times 10^{3} $ & $7.18 \times 10^7$ & $122$\\
  \hline
   B2  & $ 1 \times 10^3 $ & $ 1.522 \times 10^7$ & $ 128$ & $ $ &  F2  & $ 5 \times 10^{3} $ & $7.18 \times 10^7$ & $165$\\
  \hline
   B3  & $ 1 \times 10^3 $ & $ 1.522 \times 10^7$ & $ 169$ & $ $ &  F3  & $ 5 \times 10^{3} $ & $7.18 \times 10^7$ & $248$\\
  \hline
   B4  & $ 1 \times 10^3 $ & $ 1.522 \times 10^7$ & $ 252$ & $ $ &  F4  & $ 5 \times 10^{3} $ & $7.18 \times 10^7$ & $330$\\
  \hline
   B5  & $ 1 \times 10^3 $ & $ 1.522 \times 10^7$ & $ 333$ & $ $ & $  $ & $ $ & $ $ & $ $\\
  \hline \hline
   C1  & $ 2 \times 10^3 $ & $ 2.9 \times 10^7$ & $ 126$ & $ $ &  G1  & $ 6 \times 10^{3} $ & $8.6 \times 10^7$ & $122$\\
  \hline
   C2  & $ 2 \times 10^3 $ & $ 2.9 \times 10^7$ & $ 167$ & $ $ &  G2  & $ 6 \times 10^{3} $ & $8.6 \times 10^7$ & $165$\\
  \hline
   C3  & $ 2 \times 10^3 $ & $ 2.9 \times 10^7$ & $ 252$ & $ $ &  G3  & $ 6 \times 10^{3} $ & $8.6 \times 10^7$ & $248$\\
  \hline
   C4  & $ 2 \times 10^3 $ & $ 2.9 \times 10^7$ & $ 336$ & $ $ &  G4  & $ 6 \times 10^{3} $ & $8.6 \times 10^7$ & $330$\\
  \hline \hline
   D1  & $ 3 \times 10^3 $ & $ 4.32 \times 10^7$ & $ 124$ & $ $ &  H1  & $ 8 \times 10^{3} $ & $1.14 \times 10^8$ & $122$\\
  \hline
   D2  & $ 3 \times 10^3 $ & $ 4.32 \times 10^7$ & $ 166$ & $ $ &  H2  & $ 8 \times 10^{3} $ & $1.14 \times 10^8$ & $165$\\
  \hline
   D3  & $ 3 \times 10^3 $ & $ 4.32 \times 10^7$ & $ 249$ & $ $ &  H3  & $ 8 \times 10^{3} $ & $1.14 \times 10^8$ & $248$\\
  \hline
   D4  & $ 3 \times 10^3 $ & $ 4.32 \times 10^7$ & $ 332$ & $ $ &  H4  & $ 8 \times 10^{3} $ & $1.14 \times 10^8$ & $330$\\
  \hline \hline
\end{tabular}
\end{center}
\caption{Simulation ID, number of clusters, total mass and cut-off radius of the CC.
         Note that the table is vertically split in two subtables.
\label{tab.PowerLaw}
}
\end{table*}

Our parameter space consists of the number of clusters $\rm N$ and the
initial radius of the CC, $R_{\rm CC}$. The total mass ${M}_{\rm CC}$ of the CC
is a consequence of $N$, because the masses of the clusters are assumed to
follow a power law. The total radius that we use varies from 45~pc to 330~pc.
Given the mass and the size of the CCs, the initial escape
speeds at the centres of the CCs are between $27-137 \kms$.  All
details for all simulations are given in table \ref{tab.PowerLaw}.

\subsection{Results of the simulations}

\begin{figure*}
\resizebox{\hsize}{!}
          {\includegraphics[scale=1,clip]{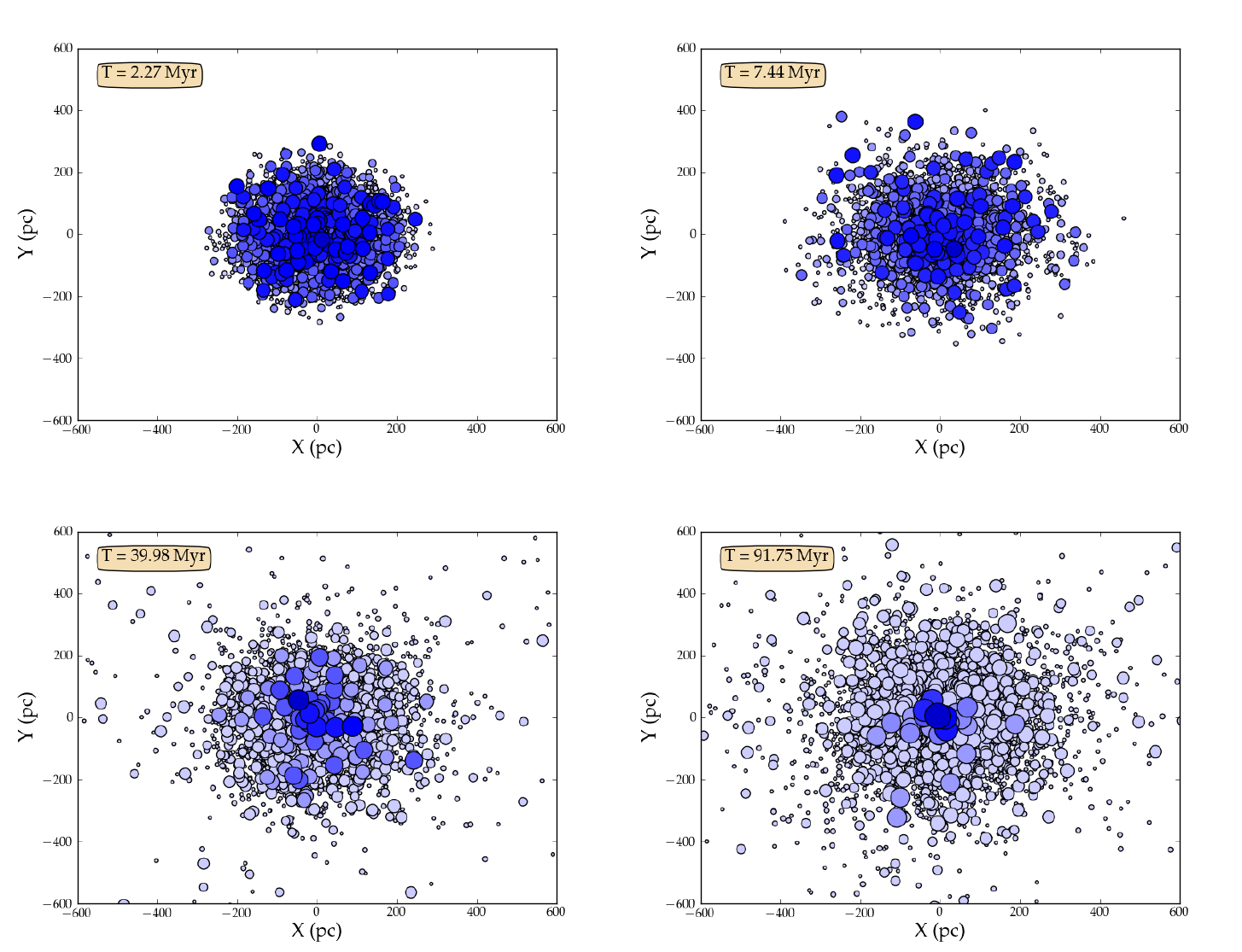}}
\caption
   {
  Formation of the UCD seed at the centre of the CC. We show a projection in the
  X--Y plane of all individual clusters for the simulation F3.  The radii of the
  clusters have been artificially magnified, heavier members have larger sizes
  and darker colours {\em relative to every panel for the sake of visibility}.
  This means that even if the colours of the heaviest clusters in the last panel
  are as dark as the most massive ones in the first panel, the clusters in the
  last panel are heavier and larger.  After 7.44 Mys we can already see how the
  more massive clusters start to agglomerate at the centre of the CC.  Later, at
  $T \sim 40$ Myr, all of them are confined to the central part of the CC and in
  the last panel we can see that only a handful of clusters are heavy and a very
  massive cluster is sitting at the very centre, while lighter clusters occupy
  all of the remaining space.  The mass of this very massive cluster is $2.9
  \times 10^6 \Msun$ and constitutes the seed of the UCD.  See
  \url{http://members.aei.mpg.de/amaro-seoane/ultra-compact-dwarf-galaxies}, model F3 for
  an animation of the process.
} \label{fig.6k_r248_rblue} \end{figure*}

\begin{figure*}
\resizebox{\hsize}{!}
          {\includegraphics[scale=1,clip]{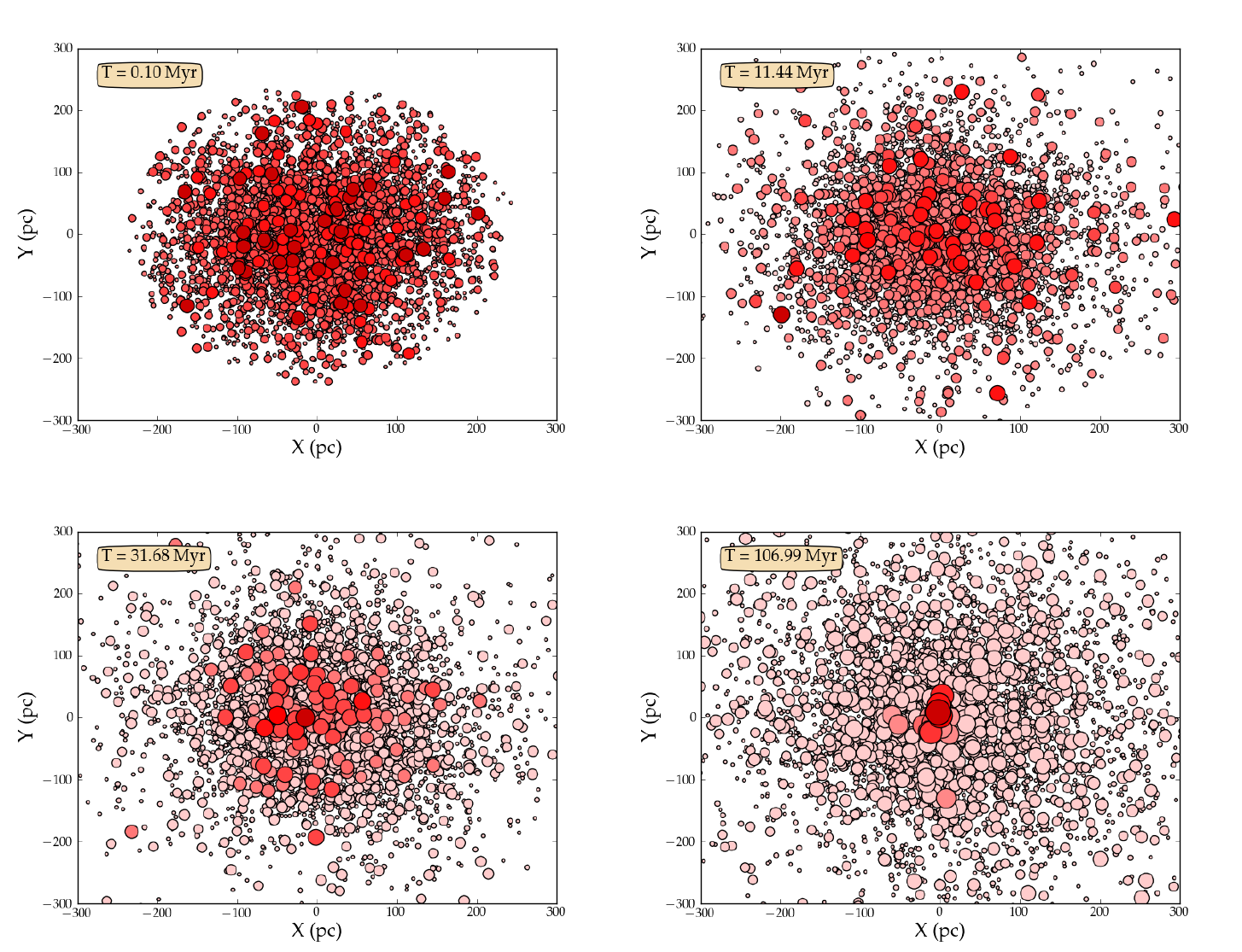}}
\caption
   {
  Same as in Fig.(\ref{fig.6k_r248_rblue}) but for simulation G3.  In this case
  we show a zoom of diameter 600 pc. As in the first figure, after some $\sim
  100\, \Myr$ we have a very massive cluster at the centre and all other clusters
  are much lighter. The heaviest cluster at this time has a mass of $5.5 \times
  10^5\, \Msun$, while clusters with masses $5.2 \times 10^5\, \Msun$, $5.0
  \times 10^5\, \Msun$, $1.9 \times 10^5\, \Msun$, $1.4 \times 10^5\, \Msun$ and
  $6.5 \times 10^4\, \Msun$ lie very close to the centre of the CC. See
  \url{http://members.aei.mpg.de/amaro-seoane/ultra-compact-dwarf-galaxies}, model G3 for a
  movie of the figure.
   }
\label{fig.5k_r248_rred}
\end{figure*}

\begin{figure}
\resizebox{\hsize}{!}
{\includegraphics[scale=0.5,clip]{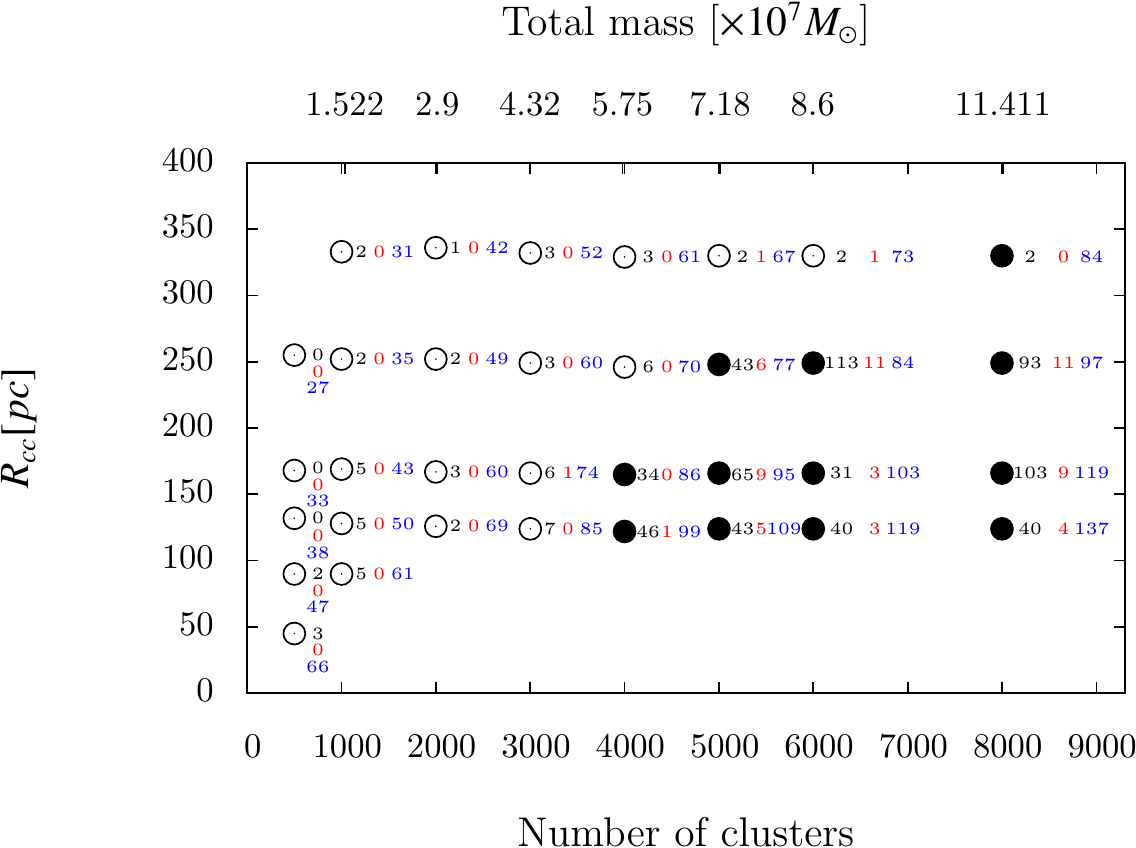}}
\caption
{
  Outcome of the CC simulations.  The x-axis shows the number of clusters in each
  simulation, while the y-axis shows the initial radius of the CC. The upper
  x-axis shows the total mass of the system in $\rm M_\odot$.  Every circle
  corresponds to a single entry of Table \ref{tab.PowerLaw} in a way such that
  the circle at the bottom left corresponds to the simulation with ID A1 and the
  circle at the top right corresponds to the simulation with ID H4.  An open
  circle indicates a simulation where the IMBH finally escaped the CC. On the
  other hand, a filled circle represents a simulation where the IMBH remained
  bound to the system. Next to every circle there are three numbers.  The first
  (black) shows the number of clusters hit by the IMBH until either it escapes the
  CC or it is captured by a cluster. The second (red) number is the number of
  stars that are tidally disrupted by the IMBH and the number of star-star
  collisions triggered by the IMBH in the clusters. The third number indicates the
  initial escape velocity at the centre of the CC in \kms.
}
\label{fig.Results}
\end{figure}

In figure \ref{fig.Results} we present the results of our 34 CC simulations.
In simulations A1-A5, B1-B5, C1-C4, D1-D4, E3-E4, F4, G4 and H4 the
IMBH escapes the CC after between zero and a few interactions with clusters. These cases
correspond to smaller-mass CCs or to low initial concentrations.
In simulations E1-E2, F1-F3, G1-G3, and H1-H4, which are more representative
of observed CCs, the IMBH is captured in the CC after
a significant number of interactions and ends up being trapped by an individual
cluster, which can be the UCD seed (cases E1, E2, F2, F3, and H2).
We show two particular cases which led to the capture of the IMBH in figures
\ref{fig.6k_r248_rblue} and \ref{fig.5k_r248_rred}.

\begin{figure}
\resizebox{\hsize}{!}
          {\includegraphics[scale=1,clip]{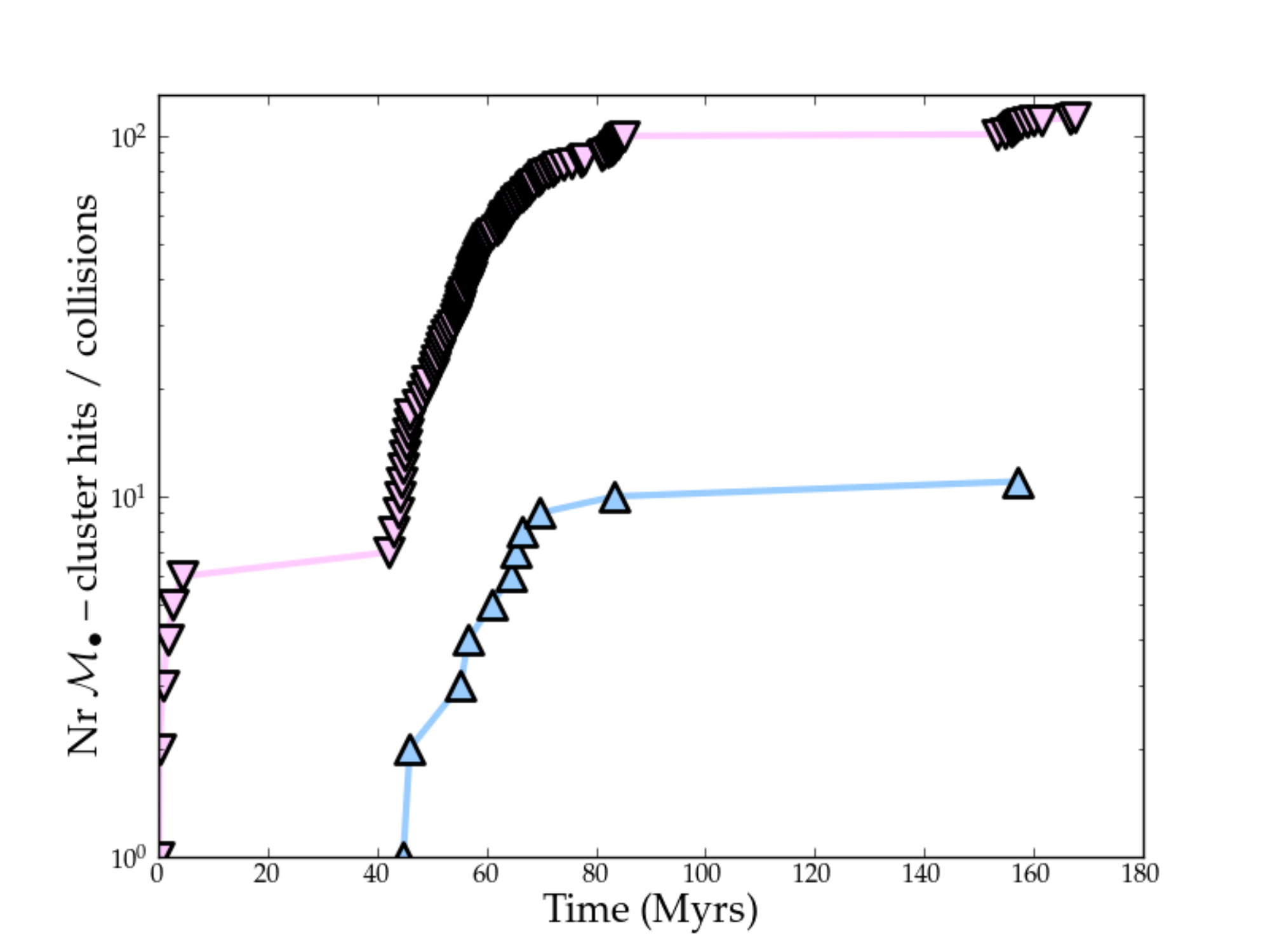}}
\caption
   {
  Cumulative number of IMBH and cluster hits for the simulation G3 (inverted,
  light magenta triangles) and of stellar collisions leading to a disruption
  (blue triangles) as a function of time.
   }
\label{fig.Cluster_Hits_G3}
\end{figure}

The IMBH goes through a very large number of interactions with individual
clusters until it is eventually trapped.  This number depends on the density of
clusters in the CC.  In 6 simulations, the IMBH gets captured by a cluster that
has not yet merged with other clusters. In 5 simulations, the cluster that
captures the IMBH is the central cluster of the CC, the seed UCD.  We show in
table \ref{tab.Table1} the details about the cluster that captures the IMBH, the
distance from the centre where this takes place, and the mass of the most
massive cluster in the system at the time of the IMBH-capture, i.e., the mass of
the UCD seed.  An interesting process in the dynamical evolution of the system
is that the IMBH triggers stellar collisions, i.e., stars are set on such an
orbit that they collide and disappear from the system. We note that only in one
case, in simulation F1, one star was torn apart by the tidal forces of the IMBH
acting on a star. The middle number next to each circle of figure
\ref{fig.Results} corresponds to star-star {\em collisions triggered by the
IMBH} in the clusters.  We can conclude that one should expect a star-star
collision in a CC every $5-8 \rm Myr$.  In figure \ref{fig.Cluster_Hits_G3} we
show the accumulated number of stellar collisions that led to a disruption in
function of the time for simulation G3, as well as the accumulated number of
hits between the IMBH and a cluster.

\begin{figure}
\resizebox{\hsize}{!}
{\includegraphics[scale=0.5,clip]{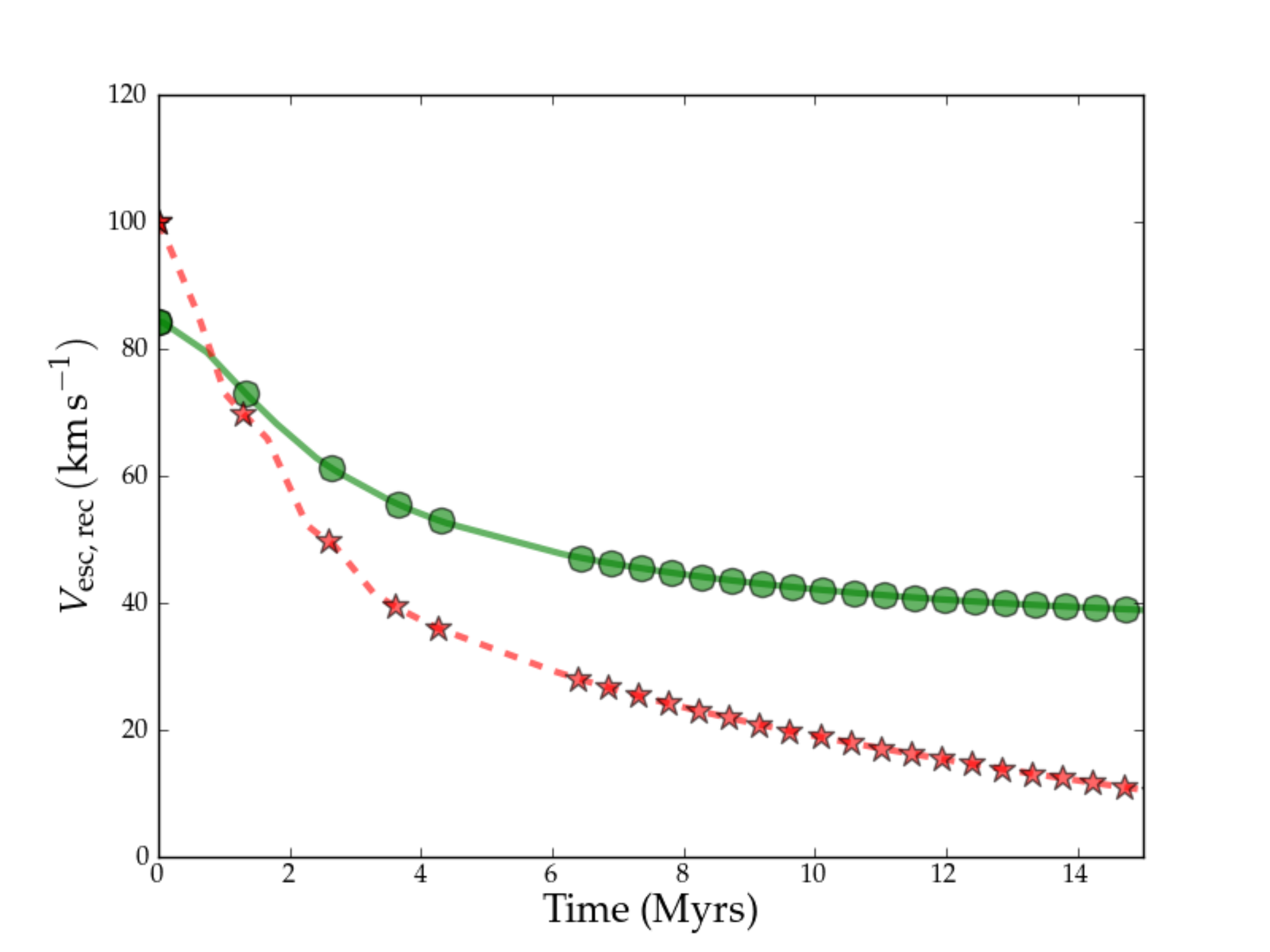}}
\caption
{
  IMBH speed (dashed, red line with stars) and instantaneous escape velocity
  (solid, green curve with spheres) for the IMBH as a function of time in
  simulation G3. Even though initially the IMBH recoiling speed is higher than the
  required threshold to escape the CC, soon after $\sim 0.80$ Myrs the
  interactions with individual clusters lower its kinetic energy and it is
  trapped in the CC, in the meaning that the speed drops below the threshold.
}
\label{fig.Velocities}
\end{figure}

The third number next to each circle of Figure \ref{fig.Results} is the initial
escape velocity at the centre of the CC. As it is obvious, CCs with values $ <
100 \rm km \, s^{-1}$ retain the IMBH due to our choice of the initial recoiling
speed. An interesting case is simulation H4 in which the escape velocity is $84
\rm km \, s^{-1}$, but the IMBH escapes because the system is initially not very
concentrated and the IMBH has only 2 interactions with clusters. In this case,
the energy of the IMBH did not decrease enough to be trapped in the CC.
Simulation G3 corresponds to the opposite situation. Even though the escape
speed is the same as in H4, the IMBH remains in the system because the CC is
denser, so that the IMBH has a chance of interacting significantly with clusters and,
hence, of decreasing its kinetic energy below the threshold. In figure
\ref{fig.Velocities} we have the evolution of the velocity of the IMBH in
simulation G3 compared with the escape velocity at the radius of the CC where
the IMBH is. Initially, the escape velocity is lower than the velocity of the
IMBH, ensuring the escape of the IMBH from the system, but the IMBH loses energy
rapidly during the first few Myr because of its interactions with clusters.

\begin{table*}
\begin{center}
\begin{tabular} {| c | c | c | c | c | c | c | c | c |}
 \hline
ID &  Coll  & T[$\rm Myr$] & $ \rm R_{\rm capt} [\rm pc]$  & $ \rm \Mcl [M_\odot] $ & $\rm {\cal M}_{\rm UCD} [M_\odot]$ & $ T_{\rm DF} [{\rm Myr}]$ & $T_{\rm FH}[\rm Myr]$ & $ T_{\rm FC}[\rm Myr]$ \\
 \hline
 \hline
 \hline
E1 & $1$ & $14$   & $9.6$  & $1.9 \times 10^6$ & $1.9 \times 10^6$ & $142$ & $0.197$ & $12.57$\\
\hline
E2 & $0$ & $38.2$ & $10.3$ & $1.2 \times 10^6$ & $1.2 \times 10^6$ & $129$ & $0.09$ & $33.59$\\
\hline
\hline
F1 & $4$ & $9.7$ & $42$ & $ 5 \times 10^3$ & $6.5 \times 10^5$ & $2400$ & $0.047$ & $2.14$\\
 \hline
F2 & $9$ & $28.2$ & $18.4$ & $7.3 \times 10^5$ & $7.3 \times 10^5$ & $240$ & $0.067$ & $0.54$\\
 \hline
F3 & $6$ & $118$ & $15.4$ & $2.9 \times 10^6$ & $ 2.9\times 10^6$ & $367$ & $0.1$ & $0.28$\\
 \hline
 \hline
G1 & $3$ & $10.14$ & $45.6$ & $2.5 \times 10^3$ & $1.1 \times 10^6$ & $4300$ & $0.011$ & $0.35$\\
 \hline
G2 & $3$ & $13.1$ & $23.8$ & $1.5 \times 10^4$ & $6.6 \times 10^5$ & $762$ & $0.009$ & $0.009$\\
 \hline
G3 & $11$ & $167.4$ & $92.7$ & $2.5 \times 10^3$ & $4 \times 10^6$ & $7900$ & $0.1$& $44.8$\\
 \hline
 \hline
H1 & $4$ & $11.7$ & $15.5$ & $5.6 \times 10^5$ & $1.3 \times 10^6$ & $26$ & $0.012$ & $3.65$\\
 \hline
H2 & $9$ & $20.1$ & $17.8$ & $1.8 \times 10^6$ & $1.8 \times 10^6$ & $360$ & $0.15$ & $5.32$\\
 \hline
H3 & $11$ & $49.9$ & $30.2$ & $1.5 \times 10^5$ & $9.7 \times 10^5$ & $167$ & $0.28$ & $9.54$\\
 \hline
\end{tabular}
\end{center}
\caption
{
  Data for the simulations where the IMBH was captured by a cluster of the CC.
  The first column shows the ID of the simulation (see table \ref{tab.PowerLaw}).
  The second column shows the number of stellar collisions triggered by the IMBH.
  The third column displays the time of capture of the IMBH by a single cluster.
  The fourth shows the distance from the centre of the CC, where the IMBH was
  captured. The next column gives us the mass of that cluster and the mass of the
  heaviest cluster in the CC by that time; i.e. the mass of the forming UCD. The
  sixth column corresponds to an estimate for the IMBH to reach the centre of the
  CC by dynamical friction (see text). The last two columns show the time the IMBH
  hits a cluster for the first time and the time of the first stellar collision
  in the CC. In the particular case of simulation F1 there was a tidal disruption
  of a star by the IMBH.
}
\label{tab.Table1}
\end{table*}

\section{Interactions between multiple recoiling IMBHs and clusters in a CC}

In this section we investigate a scenario in which $f_{\bullet} > 1$.  We use
the initial configuration of F3 as described in table \ref{tab.PowerLaw} as
our fiducial CC system and study the evolution of systems of five and ten IMBHs
at large. For this, we set them initially close to the centre and allow them to
be kicked off the host cluster at the same time, $T = 0$, as a simplifying
assumption. In real systems there will be a time lag:

\begin{equation}
\tau_{\rm bin}=\tau_{\rm run}\, \tau_{\rm IMBH} \,\tau_{\rm merg}.
\end{equation}

\noindent
where $\tau_{\rm run}$ is the timescale for a cluster to evolve to the
runaway phase, $\tau_{\rm IMBH}$ is the timescale for the VMS to become
unstable and form an IMBH and $\tau_{\rm merg}$ is the timescale for the
cluster to merge with another cluster.  The phenomena involved are various and
the assumptions inherent to $\tau_{\rm run}$ and $\tau_{\rm IMBH}$ prevent
realistic estimates, as we explained in the introduction.  On the other hand,
\cite{pau2006} estimate that $\tau_{\rm merg} \sim$ 7 Myr, which compared to
the timescale for the CC to reach the seed UCD phase, of the order of $\sim
100$ Myr, is a rather short interval of time and can be regarded as
instantaneous.  In view of these arguments, we assume that the IMBHs are expelled
instantaneously from their host clusters at different places of the CC at $T =
0$.

\begin{figure}
\resizebox{\hsize}{!}
          {\includegraphics[scale=1,clip]{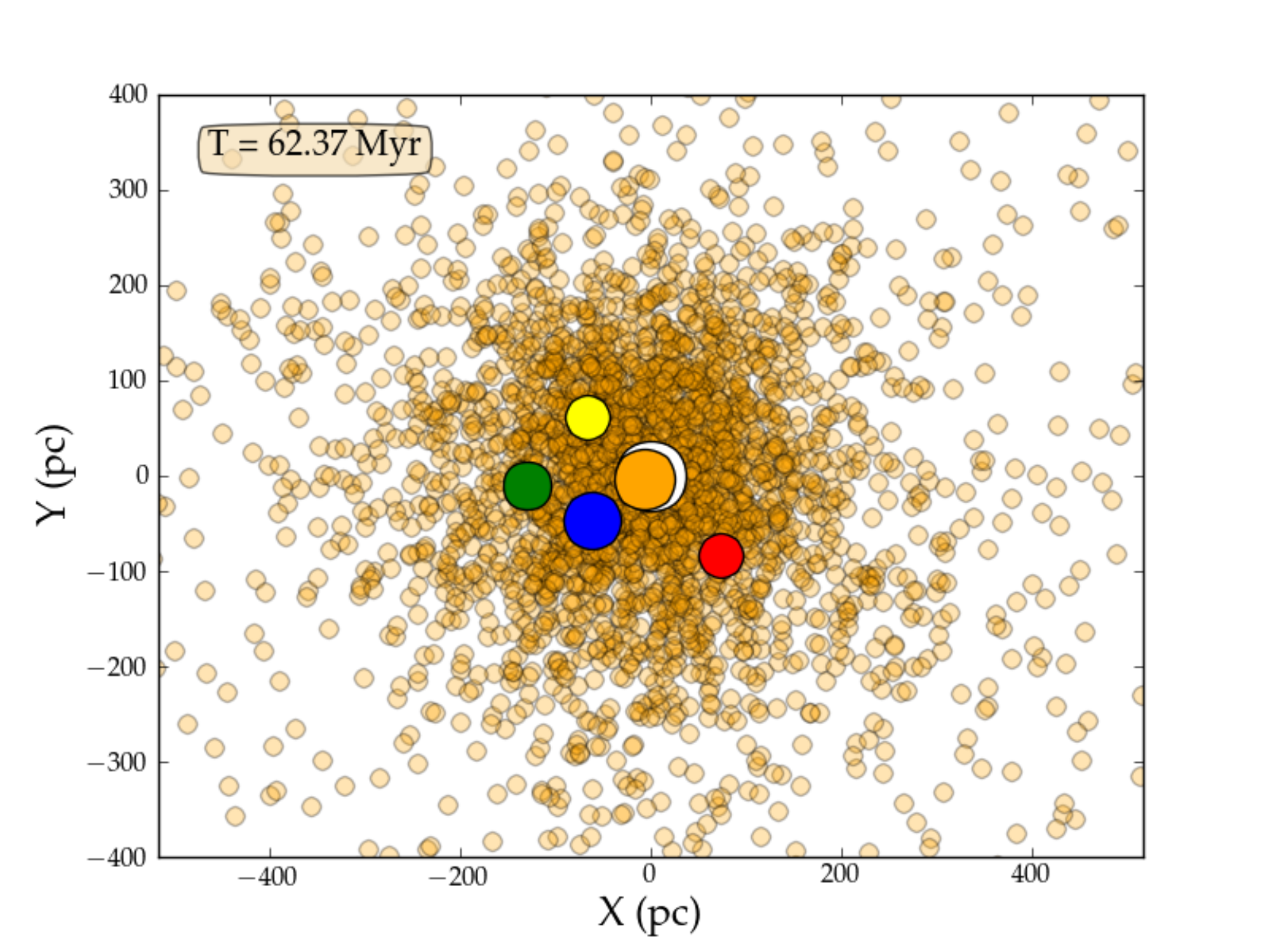}}
\caption
   {
Projection in X--Y of all clusters for the simulation in which we have initially
5 IMBHs. We show in white, orange, blue, green, yellow, and red the clusters that
captured the holes (or will capture, if in satellite, see text). For clarity we
depict all other clusters with the same radius and colour (light orange).
The green cluster harbours two IMBHs and the blue cluster too. The later one
merged with an IMBH and after that with another one which contained another IMBH.
   }
\label{fig.Data_Various_BHs}
\end{figure}

In table \ref{tab.FiveMBHs} we show the results for the first simulation, in
which we place five IMBHs around the centre, as indicated in column number
three.  IMBHs \#2 -- 5 have been distributed over the surface of a sphere of
radius 17.32 pc and only one, \#1, is very close to the centre, to avoid the
artificial formation of various binaries of IMBHs when we start the simulation.
We assign the holes initial recoil speeds between $50-100\,{\rm km\,s}^{-1}$
and different directions and then let the system evolve. We find that after
some $\sim 34$ Myr all IMBHs have been either captured by an individual cluster
which is sinking the the centre due to DF, or formed a satellite with a
cluster.  In figure \ref{fig.Data_Various_BHs} we show the CC at $T = 62.37$
Myr. We stop the simulation at that time because the satellites are consuming
all of the computational power. In the process and up to that time, there are 7
stars that have been disrupted in the CC, as we can see in the table.

\begin{table*}
\begin{center}
\begin{tabular} {| c | c | c | c | c | c | c | c | c |}
 \hline
IMBH ID &  Outcome & $\rm R_{\rm init}$ (pc) & $\rm V_{\rm init}/V_{\rm esc} $ & T (Myr) & $\rm R_{\rm fin}^{\rm UCD}$ (pc)  & $ M_{\rm cl} (M_\odot) $ & \# Stellar disr \\
 \hline
 \hline
1 & capture    & 0.0024  & 0.89 & 14.24  & 51.2 & $2.5 \times 10^3$  & 1 \\
\hline
2 & capture    & 17.32   & 0.70 & 12.41  & {103} & $7.85 \times 10^5$ & 2 \\
\hline
3 & satellite & 17.32   & 0.71 & 34     & 0       & $9.6 \times 10^5$  & 2 \\
\hline
4 & satellite & 17.32   & 0.94 & 9.45   & 9.5  & $5.44 \times 10^5$ & 1 \\
\hline
5 & capture    & 17.32   & 0.76 & 2.35   & 118.2 & $5 \times 10^5$    & 1 \\
\hline
\end{tabular}
\end{center}
\caption
{
  Data for the simulation with five IMBHs in a CC.  The first column shows the ID
  of the IMBH, the second column the outcome of the BH after 35 Myr, which can be
  either a capture, a satellite (the IMBH is orbiting a cluster and will
  eventually merge with it) or an escape (the IMBH escapes the whole CC). The
  third column displays the initial distance of the IMBH from the center of the
  CC. The fourth corresponds to the initial velocity of the  normalised to the
  local escape velocity from the cluster. The fifth gives the time at which the
  outcome was measured. The sixth shows the final distance of the capturing cluster
  from the most massive cluster of the system, the seed UCD. In this case, IMBH
  \#3 is captured by the seed, and thus this distance is zero.  The seventh is the
  mass of the capturing cluster at the time of capture. Finally, the last column
  shows the number of stellar collisions in clusters that have been triggered by
  the IMBH.
}
\label{tab.FiveMBHs}
\end{table*}

In table \ref{tab.TenMBHs} we repeat the same exercise but for a system with 10 IMBHs.
The initial setup is identical to the previous one. We find that in this case three
holes leave the system due to an increase in their kinetic energy. The rest of them
have formed  a hard binary with a cluster and will eventually be captured.

\begin{table*}
\begin{center}
\begin{tabular} {| c | c | c | c | c | c | c | c | c |}
 \hline
IMBH ID &  Outcome & $\rm R_{\rm init}$ (pc) & $\rm V_{\rm init}/V_{\rm esc} $ & T (Myr) & $\rm R_{\rm fin}^{\rm UCD}$ (pc)  & $ M_{\rm cl} (M_\odot) $ & \# Stellar disr \\
 \hline
 \hline
1  & satellite &  0.0018 & 0.56 &  1.9  & 130.7 & $7.5 \times 10^5$  & 2 \\
\hline
2  & satellite &  17.33  & 0.97 & 23.8  & 130.7 & $1.6 \times 10^6$  & 2 \\
\hline
3  & satellite &  17.33  & 0.99 &  8.9  & 112.2 & $1.44 \times 10^6$ & 3 \\
\hline
4  & satellite &  17.33  & 0.89 &  9.1  & 163.5 & $2.2 \times 10^6$  & 1 \\
\hline
5  & escaper   &  17.33  & 0.56 &  -    & - & -                  & 5 \\
\hline
6  & escaper   &  17.33  & 0.59 &  -    & - & -                  & 0 \\
\hline
7  & escaper   &  17.33  & 0.88 &  -    & - & -                  & 7 \\
\hline
8 & satellite  &  17.33  & 0.90 & 10.2  & 8.7& $2.7 \times 10^6$  & 1 \\
\hline
9  & satellite &  17.33  & 0.56 &  5.7  & 140.5& $7.75 \times 10^6$ & 0 \\
\hline
10 & satellite &  25.99  & 0.72 &  4.4  & 140.5 & $1.47 \times 10^6$ & 1\\
\hline
\end{tabular}
\end{center}
\caption
{
  Same as in table \ref{tab.FiveMBHs} but for ten holes. In this case three IMBHs
  leave the CC. We find 22 stellar disruptions during the simulation.
}
\label{tab.TenMBHs}
\end{table*}

\section{Lower kinetic energy loss}
\label{sec.5percent}

In the simulations of the previous sections we assumed a loss of relative
kinetic energy of $\sim 20\%$ for the hits that led to a fly-through, although
it could be much larger than that, as we saw in figure \ref{fig.histogram}.
While this is true for a bit less than $50\%$ of all systems, the rest of them
had a peak in the distribution around $\sim 5\%$. We have addressed the
situation of a larger loss first, because it leads to more interesting effects
from a pure dynamical standpoint.

\begin{figure}
\resizebox{\hsize}{!}
          {\includegraphics[scale=1,clip]{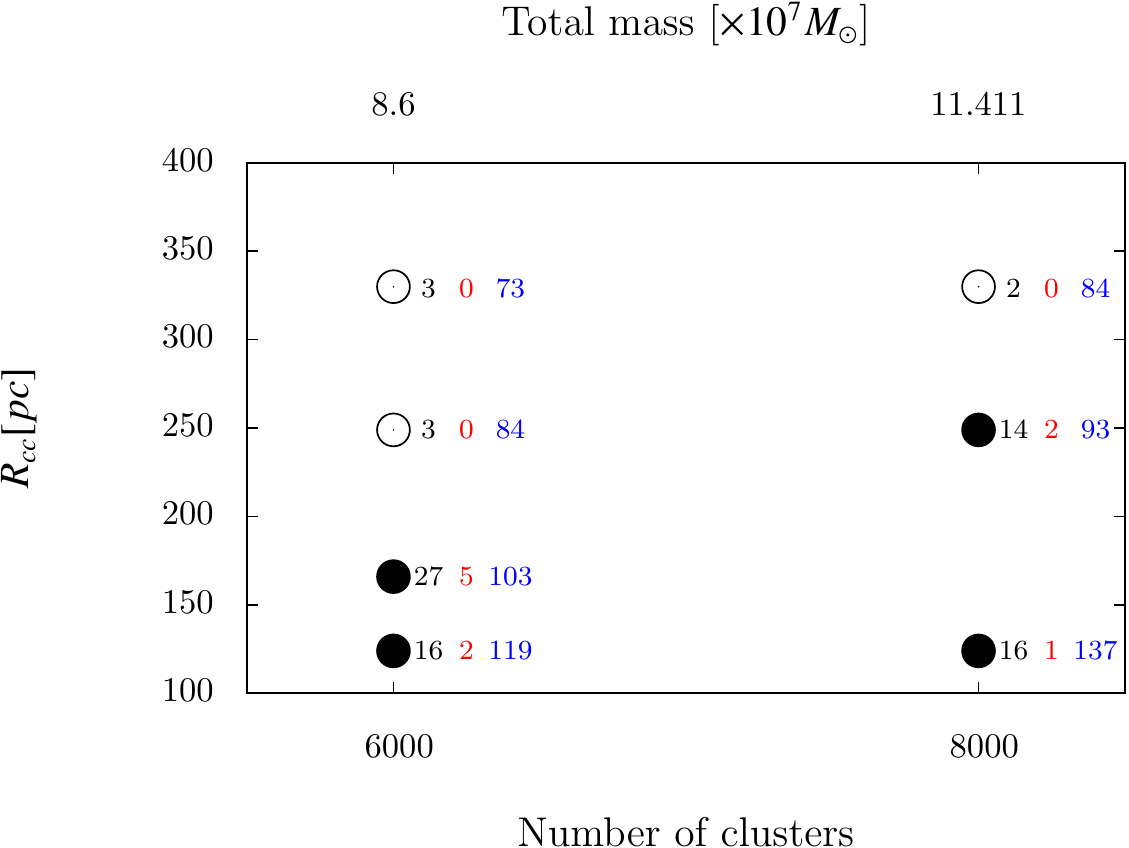}}
\caption
   {
Same as figure (\ref{fig.Results}) but assuming a fixed loss of kinetic energy
of $5\%$ after every hit for
the fly-throughs.
   }
\label{fig.r_vs_m_5percent}
\end{figure}

However, we deem it necessary to we repeat some experiments in the evolution of
the CC to understand the other regime. Therefore, we repeat experiments G1, G2,
G3, G4, H1, H3, H4 of table \ref{tab.PowerLaw} but this time we assume a loss
of $5\%$ after every hit for the fly-throughs. In figure
(\ref{fig.r_vs_m_5percent}) we can see the results.  We have reduced the
exploration to the range of radii and total mass that could be more interesting
for our analysis. We can see that although the total number of stellar
disruptions is signficantly reduced, it is not zero. Also, in four
configurations the IMBH at large is captured eventually by the forming CC.

\section{Summary and conclusions}
\label{sec.conclusions}

In this work we have presented results that address the formation of UCD from
young clusters, and the role of recoiling IMBHs in a CC. The formation of the
IMBH in clusters is used as a working hypothesis, and hence also the
possibility that these interact with the young clusters. For that, we first ran
a set of $\sim \, 200$ direct-summation $N-$body simulations that covers the
parameter space for individual IMBH--cluster encounters. We methodically varied
the mass ratio between the IMBH and the cluster, the relative velocity, and the
impact parameter. This allowed us to build a grid with the expected outcome of
the interaction and the modification of the kinetic energy of the IMBH.  Later
we ran additional direct-summation $N-$body simulations for a scenario in which
one IMBH is at large in a CC. The IMBH is assumed to be the result of the
coalescence of two holes, which led to the expulsion of the hole from the
initial host cluster.  We studied the dynamical evolution of this single IMBH
in an evolving CC. Parallel to the individual interactions between the IMBH and
clusters in the CC, which are corrected using the above-mentioned table,
clusters are colliding and merging with each other, which results in the
formation of a run-away individual cluster, which typically after $\sim 100$
Myrs contains almost all of the mass of the CC. This is what we designate ``the
seed of an ultra-compact dwarf galaxy'', since this very massive cluster is the
result of the successive amalgamation of smaller clusters in the initial
distribution of the CC.

We find that for realistic CCs (i.e. those which resemble observations, such as the
knots of the Antenn{\ae}), the IMBH is either eventually captured by the seed
UCD (in those simulations less dense initially) or by a smaller cluster (in the
simulations with the largest concentrations of clusters at the centre) which,
however, is close to the centre of the CC, so that it will in the course of
time sink down to the very centre, where the seed UCD is settled. The typical
timescale for this trapping is of about $\sim 200$ Myr.

We can see this by estimating the dynamical friction time $T_{\rm DF}$. This is
the timescale for the IMBH captured in a cluster to reach the centre. For an object
with mass $\rm m$ moving in a system of total mass $\rm M$ it is given by
\citep[see e.g.][]{BinneyTremaine08}

\begin{equation} \label{t_df}
T_{\rm DF} = \frac{1.17}{\ln{\Lambda}} \frac{\rm M}{\rm m} \frac{\rm r}{\rm V_h},
\end{equation}

\noindent
where $\rm r$ is the distance from the centre of the system, ${\rm V}_h$ is the
root mean square (RMS) velocity dispersion of the system and $\ln{\Lambda}$ the
Coulomb logarithm, which is of the order of unity. From table \ref{tab.Table1}
we can see that in almost half of the cases in which the IMBH was retained in
the CC, it is captured by the most massive cluster of the system, the seed UCD.
$T_{\rm DF}$ is in all cases a few tens or hundreds of Myrs.  On the other
hand, when the IMBH gets captured by a smaller cluster (6 out of the 11
simulations), $T_{\rm DF}$ is of the order of $\sim 1$ Gyr, still well below a
Hubble time.  We note also that this analytical calculation is an
overestimate, because the CC evolves dynamically with time and there is a
huge accumulation of mass in the innermost region which will significantly
reduce the timescale for the IMBH to reach the seed UCD.

When the IMBH remains bound to the CC, the average time for it to hit a cluster
is $0.16-0.43 \rm Myr$. On the other hand, the mean time taken by the IMBH to
fly through a star cluster is of the order of $0.1 \rm Myr$. Hence, after
recoiling and before getting captured, the IMBH spends $1/3$ of its time
interacting with clusters, so that the possibility to find an IMBH in a cluster
of a newly formed (less than $100 \rm Myr$ old) CC is about $\sim 30\%$

We repeated the exercise with a CC harbouring initially 5 and 10 IMBHs which
were distributed with different velocities. We find that after some $\lesssim
30$ Myr most of the holes are either captured by a single cluster or have
formed a hard binary with one in regions relatively close to the runaway
cluster, the seed of a UCD which is forming in the CC. We cannot follow the
further evolution of the system due to the limitations inherent to our
approach.  We also note that gas is very likely to play an important role in
the whole process.  In particular, in some CCs the oldest cluster is located at
the centre of the gas cloud \citep{WhitmoreEtAl10}. In our simulations we have
neglected this, since we are limited by our codes, which rely on pure particle
dynamics. Still, even if we could actually have implemented a (rough) approach
for the gas with an external force, the complexity of the
problem justifies our first approach. We have decided to postpone the role of
the gas for upcoming work. The same applies to mass loss because of stellar evolution,
although statistically, since the IMBH interacts with clusters of different masses,
the global dynamical evolution is well represented by our models, within our
limitations.

Also, reducing the relative kinetic energy loss for fly-throughs leads to a reduced
number of tidal disruption events, but we still find some systems for which the
implications are similar to the analysis that used a larger loss.

While the number fraction of IMBH in the mass-range of $10^{2-4}\,M_{\odot}$ in
CCs is an unknown, they sink to the centre in a time which is much shorter than
the Hubble time. The scenario that we have described here leads to the
formation of a very massive black hole at the centre of the UCD, with a mass
that depends on unknowns, such as the formation rate of IMBHs in the CC.
The internal velocities of the systems we study are not as extreme as those
explored by \cite{MerrittSchnittmanKomossa09} in the context of
hypercompact stellar systems, because the seed UCD inherits the central
velocity from the resulting mergers between individual clusters. When the UCD
is formed, the velocity will roughly be what one can expect from a dense
stellar system in dynamical equilibrium.  A very interesting feature of the
process of sowing an UCD with an IMBH is that independently of whether the IMBH
stays in the CC or escapes, it triggers star-star collisional disruptions in
the clusters it hits. This could be envisaged as an electromagnetic signature
of the scenario.

\acknowledgments

It is a pleasure for PAS to thank Rainer Sch{\"o}del and Emilio Alfaro for
their invitation to the IAA, where some of the work was done. He is
particularly indebted to Rainer and Elena Navajas for their support and for the
psetta maxima. This work was supported in  part by the National Science
Foundation under Grant No.\ 1066293 and the hospitality of the Aspen Center for
Physics.  The work of SK was funded by the German Science Foundation (DFG) via
SFB/TR7 on ``Gravitational Waves'' and the the German Academic Excange Service
(DAAD). MCM was supported in part by NASA ATP grants NNX08AH29G and NNX12AG29G.  FAR
acknowledges support from NSF Grant PHY-0855592 and NASA Grant NNX08AG66G.  The
use of the GRAPE-6 system of the University of Thessaloniki was supported by a
grant by the Empirikion Foundation. We acknowledge the usage of the {\sc
Tuffstein} GRAPE cluster at the AEI.

\end{document}